\documentclass[a4,11pt]{article}

\usepackage{graphicx}
\usepackage{amsmath,amsthm}
\usepackage{amssymb}
\usepackage{amsfonts}
\usepackage{latexsym}
\usepackage{comment}
\makeatletter
 
  \@addtoreset{equation}{section}
 \makeatother
\begin{document}
\title{A General Framework for the Benchmark pricing\\ in a Fully Collateralized Market~\footnote{
This research is supported by CARF (Center for Advanced Research in Finance)
and JSPS KAKENHI (Grant Number 25380389).
The authors are not responsible or liable in any manner for any losses and/or damages caused by the use of any contents in this research.
Initially, it was titled as {\it Choice of collateral currency updated}.
}  \\
}

\author{Masaaki Fujii\footnote{Faculty of Economics, The University of Tokyo: mfujii@e.u-tokyo.ac.jp},
Akihiko Takahashi\footnote{Faculty of Economics, The University of Tokyo: akihikot@e.u-tokyo.ac.jp, corresponding author}
}
\date{
7 September, 2015\\
}
\maketitle



\newtheorem{definition}{Definition}[section]
\newtheorem{assumption}{$[$ A}[section]
\newtheorem{condition}{$[$ C}
\newtheorem{lemma}{Lemma}[section]
\newtheorem{proposition}{Proposition}[section]
\newtheorem{theorem}{Theorem}[section]
\newtheorem{remark}{Remark}[section]
\newtheorem{example}{Example}[section]
\newtheorem{corollary}{Corollary}[section]
\def\n{{\bf n}}
\def\A{{\bf A}}
\def\B{{\bf B}}
\def\C{{\bf C}}
\def\D{{\bf D}}
\def\E{{\bf E}}
\def\F{{\bf F}}
\def\G{{\bf G}}
\def\H{{\bf H}}
\def\I{{\bf I}}
\def\J{{\bf J}}
\def\K{{\bf K}}
\def\L{{\bf L}}
\def\M{{\bf M}}
\def\N{{\bf N}}
\def\O{{\bf O}}
\def\P{{\bf P}}
\def\Q{{\bf Q}}
\def\R{{\bf R}}
\def\S{{\bf S}}
\def\T{{\bf T}}
\def\U{{\bf U}}
\def\V{{\bf V}}
\def\W{{\bf W}}
\def\X{{\bf X}}
\def\Y{{\bf Y}}
\def\Z{{\bf Z}}
\def\cala{{\cal A}}
\def\calb{{\cal B}}
\def\calc{{\cal C}}
\def\cald{{\cal D}}
\def\cale{{\cal E}}
\def\calf{{\cal F}}
\def\calg{{\cal G}}
\def\calh{{\cal H}}
\def\cali{{\cal I}}
\def\calj{{\cal J}}
\def\calk{{\cal K}}
\def\call{{\cal L}}
\def\calm{{\cal M}}
\def\caln{{\cal N}}
\def\calo{{\cal O}}
\def\calp{{\cal P}}
\def\calq{{\cal Q}}
\def\calr{{\cal R}}
\def\cals{{\cal S}}
\def\calt{{\cal T}}
\def\calu{{\cal U}}
\def\calv{{\cal V}}
\def\calw{{\cal W}}
\def\calx{{\cal X}}
\def\caly{{\cal Y}}
\def\calz{{\cal Z}}
%
\def\sskip{\hspace{0.5cm}}
\def\simleq{ \raisebox{-.7ex}{\em $\stackrel{{\textstyle <}}{\sim}$} }
\def\leqsim{ \raisebox{-.7ex}{\em $\stackrel{{\textstyle <}}{\sim}$} }
\def\ep{\epsilon}
\def\half{\frac{1}{2}}
\def\iku{\rightarrow}
\def\Iku{\Rightarrow}
\def\ikup{\rightarrow^{p}}
\def\inclusion{\hookrightarrow}
\def\cadlag{c\`adl\`ag\ }
\def\up{\uparrow}
\def\down{\downarrow}
\def\doti{\Leftrightarrow}
\def\douti{\Leftrightarrow}
\def\dochi{\Leftrightarrow}
\def\douchi{\Leftrightarrow}%
\def\yy{\\ && \nonumber \\}
\def\y{\vspace*{3mm}\\}
\def\nn{\nonumber}
\def\be{\begin{equation}}
\def\ee{\end{equation}}
\def\bea{\begin{eqnarray}}
\def\eea{\end{eqnarray}}
\def\beas{\begin{eqnarray*}}
\def\eeas{\end{eqnarray*}}
%
\def\hd{\hat{D}}
\def\hv{\hat{V}}
\def\hsd{{\hat{d}}}
\def\hx{\hat{X}}
\def\hsx{\hat{x}}
\def\bsx{\bar{x}}
\def\bsd{{\bar{d}}}
\def\bx{\bar{X}}
\def\ba{\bar{A}}
\def\bb{\bar{B}}
\def\bc{\bar{C}}
\def\bv{\bar{V}}
\def\balpha{\bar{\alpha}}
\def\bbalpha{\bar{\bar{\alpha}}}
\def\combi{\l(\begin{array}{c}\alpha\\ \beta \end{array}\r)}
\def\f{^{(1)}}
\def\s{^{(2)}}
\def\ss{^{(2)*}}
\def\l{\left}
\def\r{\right}
\def\a{\alpha}
\def\b{\beta}
\def\L{\Lambda}

\def\E{{\bf E}}
\def\P{{\bf P}}
\def\Q{{\bf Q}}
\def\R{{\bf R}}
\def\mbb{\mathbb}
\def\calf{{\cal F}}
\def\calp{{\cal P}}
\def\calq{{\cal Q}}

\def\ep{\epsilon}
\def\part{\partial}
\def\del{\delta}
\def\Del{\Delta}
\def\del{\delta}
\def\bull{$\bullet$}
\def\wt{\widetilde}

\def\yy{\\ && \nonumber \\}
\def\y{\vspace*{3mm}\\}
\def\nn{\nonumber}
\def\be{\begin{equation}}
\def\ee{\end{equation}}
\def\bea{\begin{eqnarray}}
\def\eea{\end{eqnarray}}
\def\beas{\begin{eqnarray*}}
\def\eeas{\end{eqnarray*}}

\def\l{\left}
\def\r{\right}

\vspace{20mm}
\begin{abstract}
Collateralization with daily margining has become a new standard in the post-crisis market.
Although there appeared vast literature on a so-called multi-curve framework, a complete picture 
of a multi-currency setup with cross-currency basis
can be rarely found since our initial attempts.
This work gives its extension regarding a general
framework of interest rates in a fully collateralized market.
It gives a new formulation of the currency funding spread which is 
better suited for the  general dependence.
In the last half, it develops a discretization of the HJM framework
with a fixed {\it tenor structure}, which makes it implementable as a traditional
Market Model.

\end{abstract}
\vspace{10mm}

\newpage
\section{Introduction}
The landscape of derivative modeling has experienced a radical change since the financial crisis in 2008.
On the one hand, evaluation of the counterparty credit risk has become an unavoidable element for all the types 
of financial contracts. This is a rather natural consequence of significant number of credit events,
well exemplified by a collapse of Lehman brothers, which was one of the most prestigious investment banks at that time.
On the other hand, clean or benchmark pricing framework has also changed significantly.
Here, explosion of various basis spreads, which were mostly negligible before the crisis, 
 and the recognition of collateral cost have played the central role.
After some initial attempts to explain the effects of collateralization, such as 
Fujii, Shimada and Takahashi (2010a)~\cite{note-curve}, Bianchetti (2010)~\cite{bianchetti} and Piterbarg 
(2010)~\cite{piterbarg},
there appeared vast literature to build a dynamic multi-curve framework such as
Mercurio (2009)~\cite{Mercurio}, Fujii, Shimada and Takahashi (2011)~\cite{dynamic_basis}, 
Cr\'epey, Grbac and Nguyen (2012)~\cite{Crepey-Grbac},
Filipovi\'c and Trolle (2013)~\cite{trolle}, Grbac, Papapantoleon, Schoenmakers and Skovmond (2014)~\cite{Grbac-Papapantoleon},
and articles in a recent book Bianchetti and Morini (editors) (2013)~\cite{Bianchetti-Riskbook} just name a few.
See Brigo, Morini and Pllavicini (2013)~\cite{Brigo-book}, Cr\'epey and Bielecki (2014)~\cite{Crepey-book}
and references therein for closely related topics, such as CVA and FVA.

A term structure model in a multi-currency setup with non-zero cross currency basis 
was firstly developed in \cite{dynamic_basis}, 
where LIBOR-OIS spreads were stochastic but currency funding spreads were left deterministic. 
The funding spreads were made stochastic firstly in Fujii, Shimada  and Takahashi (2010b)~\cite{ccc-old}
and Fujii and Takahashi (2011)~\cite{ccc-orig} in a continuous Heath-Jarrow-Morton (Heath, Jarrow and Morton, 
1992~\cite{HJM}) 
framework.

This article is a revised  and extended version of our previous works
and presents a general framework for the benchmark pricing in a fully collateralized market.
The formulation of the dynamics of the currency funding spread is changed
from the original one~\cite{ccc-old, ccc-orig}, which will be more useful 
to implement in the presence of non-zero correlation among the collateral rates and 
funding spreads. 
In particular, in the last half of the paper, we provide a complete picture of 
a discretized HJM framework for stochastic collateral rates, LIBORs, foreign exchange rates
currency funding spreads, and equities with a fixed {\it tenor structure},
which is readily implementable as a traditional Market Model (Brace, Gatarek and Musiela, 1997~\cite{BGM}).
We do not repeat the details of the curve bootstrapping procedures. In fact the 
methodology given in the previous papers~\cite{note-curve, ccc-orig, RiskBook}
can be applied without significant change.

\section{Pricing in a fully collateralized market}
Let us start from the review of \cite{note-curve, RiskBook}.
We make the following assumptions that has become popular for the benchmark pricing:
\bea
1. &&\mbox{Full collateralization (zero threshold) by cash.}\nonumber \\
2. &&\mbox{The collateral is adjusted continuously with zero minimum 
transfer amount.}\nonumber
\eea
This means that the party who has negative mark-to-market posts the equal amount of cash collateral 
to the counterparty, and this is done continuously until the contract expires.
Actually, daily margin call is becoming popular, and that is a market standard at least 
among major broker-dealers and central counterparties. This observation allows us to see the above assumptions a reasonable 
proxy for the reality. 
One might consider there is no counterparty risk remains under the above assumption.
In fact, however,  this is not always the case, if there exists a sudden jump of 
the underlying asset and/or the collateral values at the time of counterparty default.
This is the so-called ``gap risk". If this is the case, the remaining risk should be taken into account 
as a part of credit risk valuation adjustment (CVA).  
In this article, we assume that
no counterparty risk remains and focus on the clean benchmark pricing.
For the interested readers, we refer to \cite{Brigo-book,Crepey-book} as well as
Fujii and Takahashi (2013a)~\cite{imperfect_collateral} to handle more generic 
situations with non-zero credit as well as funding risks. 

Let us mention the standing assumption of the paper:\\
$\bold{Assumption}$ \\
{\it
Throughout the paper, we assume that the 
appropriate regularity conditions are satisfied whenever they are required.
In particular every local martingale is assumed to be a true martingale. 
}\\
We consider a derivative whose payoff at time $T$ is given by $h^{(i)}(T)$ in terms of currency $(i)$.
We suppose that currency $(j)$ is used as the collateral for the contract.
Let us introduce an important spread process:
\be
\label{eq-y}
y^{(j)}(t)= r^{(j)}(t)-c^{(j)}(t)~,
\ee
where $r^{(j)}$ and $c^{(j)}$ denote the risk-free interest rate and the collateral rate of the currency $(j)$, respectively.
A common practice in the market is to set $c^{(j)}$ as the overnight (ON) rate of currency $(j)$. 
Economically, the spread $y^{(j)}$ 
can be interpreted as the dividend yield from the collateral account of currency $(j)$ from the view point of 
a collateral receiver.  On the other hand, from the view point of a collateral payer, 
it can be considered as a collateral funding cost. Of course, the return from risky investments, or 
the borrowing cost from the external market can be quite different from the risk-free rate.
However, if one wants to treat these effects directly, an explicit modeling of the associated risks
is required. Here, we use the risk-free rate as net return/cost after hedging these risks,
which can be justified in a simple credit model (see the arguments given in Section 9.4.1 of \cite{Brigo-book}.).
As we shall see, the final formula does not require any knowledge of the risk-free rate, and hence
there is no need of its estimation, which is crucial for the practical implementation.

Now, we explain the derivation of the pricing formula.
If we denote the present value of the derivative at time $t$ by $h^{(i)}(t)$ in terms of currency $(i)$,
collateral amount of currency $(j)$ posted from the counterparty is given by 
\be
\frac{h^{(i)}(t)}{f_x^{(i,j)}(t)},
\ee
where $f_x^{(i,j)}(t)$ is the foreign exchange rate at time $t$ representing the price of the unit amount of currency $(j)$
in terms of currency $(i)$.  It should be interpreted that the investor posts the collateral to the counterparty 
when $h^{(i)}(t)$ is negative.

These considerations lead to the following calculation for the
collateralized derivative price, 
\bea
h^{(i)}(t)&=&\mbb{E}_t^{\mbb{Q}^{(i)}}\left[e^{-\int_t^T r^{(i)}(s)ds}h^{(i)}(T)\right] \nn \\
&&\qquad +f_x^{(i,j)}(t)\mbb{E}_t^{\mbb{Q}^{(j)}}\left[ \int_t^T e^{-\int_t^s r^{(j)}(u)du} y^{(j)}(s)\left(\frac{h^{(i)}(s)}{f_x^{(i,j)}(s)}\right)ds\right]~
\label{price_decomp}
\eea
where $\mbb{E}_t^{\mbb{Q}^{(i)}}[\cdot]$ is the time $t$ conditional expectation under the risk-neutral measure of currency $(i)$, where
the money-market account of currency $(i)$ 
\be
\beta^{(i)}_\cdot=\exp\left(\int_0^\cdot r^{(i)}_s ds\right)
\ee
is used as the numeraire. Here, the second line of (\ref{price_decomp}) represents 
the effective dividend yield from the collateral account, or the cost of posting collateral to the counterparty.
One can see the second term changes its sign properly according to the value of the contract.
An economic discussion on the {\it risk-free} rate will be given also in Section~\ref{sec-risk-free}.

By changing the measure using the Radon-Nikodym density
\be
\left.\frac{d\mbb{Q}^{(i)}}{d\mbb{Q}^{(j)}}\right|_t=\frac{\beta^{(i)}_t f_x^{(i,j)}(0)}{\beta^{(j)}_t f_x^{(i,j)}(t)}
\ee
one can show
\bea
h^{(i)}(t)=\mbb{E}_t^{\mbb{Q}^{(i)}}\left[e^{-\int_t^T r^{(i)}(s) ds} h^{(i)}(T)+
\int_t^T e^{-\int_t^s r^{(i)}(u) du}y^{(j)}(s)h^{(i)}(s)ds\right]~.
\eea
Thus, it is easy to check that the process $X=\{X_t;~t\geq 0\}$
\be
X(t):=e^{-\int_0^t r^{(i)}(s)ds} h^{(i)}(t)+\int_0^t e^{-\int_0^s r^{(i)}(u)du}y^{(j)}(s)h^{(i)}(s)ds
\ee
is a $\mbb{Q}^{(i)}$-martingale under the appropriate integrability conditions.
This tells us that the process of the option price can be written as
\be
dh^{(i)}(t)=\left(r^{(i)}(t)-y^{(j)}(t)\right)h^{(i)}(t)dt + dM(t)~
\ee
with some $\mbb{Q}^{(i)}$-martingale $M$. It can be expressed as a linear backward stochastic 
differential equation:
\bea
h^{(i)}(t)=h^{(i)}(T)-\int_t^T \Bigl(r^{(i)}(s)-y^{(j)}(s)\Bigr)h^{(i)}(s)ds-\int_t^T dM(s)~.
\eea
As a result, we have the following theorem:
\begin{theorem}
Suppose that $h^{(i)}(T)$ is a derivative's payoff at time $T$ in terms of currency $(i)$
and that currency $(j)$ is used as the collateral for the contract.
Then, the value of the derivative at time $t$, $h^{(i)}(t)$ is given by 
\bea
\label{f_pricing_formula}
h^{(i)}(t)&=&\mbb{E}_t^{\mbb{Q}^{(i)}}\left[e^{-\int_t^T r^{(i)}(s)ds}\left(e^{\int_t^T y^{(j)}(s)ds}\right)h^{(i)}(T)\right]\\
&=&\mbb{E}_t^{\mbb{Q}^{(i)}}\left[e^{-\int_t^T c^{(i)}(s)ds-\int_t^T y^{(i,j)}(s)ds}h^{(i)}(T)\right]\nn\\
&=&D^{(i)}(t,T)\mbb{E}_{t}^{\mbb{T}^{(i)}}\left[
e^{-\int_t^T y^{(i,j)}(s)ds}h^{(i)}(T)\right]~,\label{f_pricing_formula2}
\eea
where 
\be
y^{(i,j)}(s)=y^{(i)}(s)-y^{(j)}(s)~
\ee
with $y^{(i)}(s)=r^{(i)}(s)-c^{(i)}(s)$ and $y^{(j)}(s) = r^{(j)}(s)-c^{(j)}(s)$.
Here, we have defined
the collateralized zero-coupon bond of currency $(i)$ as
\be
\label{eq-D}
D^{(i)}(t,T)=\mbb{E}_t^{\mbb{Q}^{(i)}}\left[e^{-\int_t^T c^{(i)}(s) ds} \right]~.
\ee
We have also defined the ``collateralized forward measure" $\mbb{T}^{(i)}$ of currency $(i)$, for which 
$\mbb{E}_{t}^{\mbb{T}^{(i)}}[\cdot]$ denotes the time $t$ conditional expectation.
Here, $D^{(i)}(t,T)$ is used as its numeraire, and the associated Radon-Nikodym density is defined by
\be
\label{eq-Dmeasure}
\left.\frac{d \mbb{T}^{(i)}}{d \mbb{Q}^{(i)}}\right|_t = \frac{D^{(i)}(t,T)}{\beta_c^{(i)}(t)D^{(i)}(0,T)}
\ee
where 
\be
\beta_c^{(i)}(t):= \exp\left(\int_0^t c^{(i)}_s ds\right)~.
\ee
\label{theorem1}
\end{theorem}
Notice that the collateralized zero-coupon bond is actually a ``dividend yielding"
asset due to the cash flow arising from the collateral account.
This makes $D^{(i)}(t,T)/\beta^{(i)}(t)$ inappropriate for the definition 
of the forward measure. In fact it is not a $\mbb{Q}^{(i)}$-martingale 
unless one has to compensate the dividend yield $y^{(i)}$. 
This adjustment has changed $\beta^{(i)}$ to $\beta^{(i)}_c$ in the expression of the Radon-Nikodym density.
See Section~\ref{sec-disc-HJM} for more discussions on the arbitrage-free conditions
in a collateralized market where every contract has a common dividend yield $y$.

Since $y^{(i)}$ and $y^{(j)}$ denote the collateral funding cost of currency $(i)$ and $(j)$ respectively,
$y^{(i,j)}$ represents the difference of the funding cost between the two currencies.
As a corollary of the theorem, we have
\bea
\label{col-pricing-common}
h^{(i)}(t)=\mbb{E}_t^{\mbb{Q}^{(i)}}\left[e^{-\int_t^T c^{(i)}(s)ds}h^{(i)}(T)\right]=D^{(i)}(t,T)\mbb{E}_t^{\mbb{T}^{(i)}}[h^{(i)}(T)]
\eea
when both of the payment and collateralization are done in a common currency $(i)$. 

Theorem~\ref{theorem1} provides the generic pricing formula under the full collateralization 
including the case of foreign collateral currency. The result will be used repeatedly 
throughout the article as the foundation of theoretical discussions.
We would like to emphasize the
importance of the correct understanding for the reason why the collateral rate appears in 
the discounting. It is the well-known fact that the ``effective" discounting rate of an asset with 
dividend yield $y$ is given by $(r-y)$, or the difference of the risk-free rate and the dividend yield.
Under the full collateralization, we can interpret the return (which can be negative) from the 
collateral account $y=r-c$ as the dividend yield, which leads to $r-(r-c)=c$
as the discounting rate. The arguments can be easily extended to the case of foreign collateral currency.
Now, it is clear that the so called ``OIS-discounting" can be justified only when the collateral rate``$c$" is given by the 
overnight rate of the domestic currency. For example, suppose that there is a trade where the collateral rate
or ``fee" on the posted collateral, specified in the contractual agreement, is given by the LIBOR,
then one should use the LIBOR as the discounting rate of the contract, no matter how different it is from the risk-free rate.

\section{Heath-Jarrow-Morton framework under collateralization}
\label{sec-HJM}
In this section, we discuss the way to give no-arbitrage dynamics to the relevant quantities using
Heath-Jarrow-Morton framework. This is a generalized version of 
 Fujii, Shimada and Takahashi (2011)~\cite{dynamic_basis}
and Fujii and Takahashi (2011)~\cite{ccc-orig} and also provides more rigorous arguments.
\subsection{Dynamics of the forward collateral rate}
\label{sec-HJM-ois}
We work in the filtered probability space $(\Omega,\calf_t, \mbb{F}(:=(\calf_t)_{t\geq 0}), ~\mbb{P})$
where $\mbb{F}$ is an augmented filtration generated by a $d$-dimensional Brownian motion $W$.
We assume that the market is complete and that the measure $\mbb{Q}^{(i)}$
is equivalent to $\mbb{P}$. We use a simple notation $\mbb{E}_t[\cdot]$
instead of $\mbb{E}_t[\cdot|\calf_t]$ since $\mbb{F}$ is the only relevant filtration.

Firstly, we consider the dynamics of the forward collateral rate, which is defined by
\bea
c^{(i)}(t,T)=-\frac{\partial}{\partial T}\ln D^{(i)}(t,T)
\eea
or equivalently
\bea
D^{(i)}(t,T)=\exp\left(-\int_t^T c^{(i)}(t,s)ds\right)~.
\eea
Since it can be written as
\bea
c^{(i)}(t,T)=\frac{1}{D^{(i)}(t,T)}\mbb{E}_t^{\mbb{Q}^{(i)}}\Bigl[e^{-\int_t^T c^{(i)}(s)ds}c^{(i)}(T)\Bigr]
\eea
one can see  $c^{(i)}(t,t)=c^{(i)}(t)$ by passing the limit $T\downarrow t$.

Suppose that the dynamics of the forward collateral rate under the measure $Q^{(i)}$ is given by
\bea
dc^{(i)}(t,s)=\alpha^{(i)}(t,s)dt+\sigma_c^{(i)}(t,s)\cdot dW_t^{\mbb{Q}^{(i)}}~,
\eea
where $W^{\mbb{Q}^{(i)}}$ is a $d$-dimensional $\mbb{Q}^{(i)}$-Brownian motion,
$\alpha^{(i)}:\Omega\times \mbb{R}_{+} \times \mbb{R}_{+} \rightarrow \mbb{R}$
and and $\sigma_c^{(i)}:\Omega\times \mbb{R}_{+} \times \mbb{R}_{+} \rightarrow \mbb{R}^d$ 
and $\alpha(\cdot,s), \sigma_c^{(i)}(\cdot,s)$ are both $\mbb{F}$-adapted processes for every $s\geq 0$.
Here, we have used the abbreviation 
\be
\sigma_c^{(i)}(t,s)\cdot dW_t^{\mbb{Q}^{(i)}}:=\sum_{k=1}^d \Bigl(\sigma_c^{(i)}(t,s)\Bigr)_k dW^{\mbb{Q}^{(i)}}_k(t)
\ee
to lighten the notation.

\begin{proposition}
\label{prop-cfwd}
Under the setup given in Section 3.1, the no-arbitrage dynamics of the 
collateral forward rate $\{c^{(i)}(t,s), t\in[0,s]\}$ for every $s\geq 0$ is given by
\bea
&&dc^{(i)}(t,s)=\sigma_c^{(i)}(t,s)\cdot\left(\int_t^s \sigma_c^{(i)}(t,u)du\right)dt+\sigma_c^{(i)}(t,s)\cdot dW_t^{\mbb{Q}^{(i)}} \\
&&c^{(i)}(t,t)=c^{(i)}(t)~.
\eea
\begin{proof}
Simple application of It\^o's formula yields
\bea
dD^{(i)}(t,T)/D^{(i)}(t,T)&=&\left\{c^{(i)}(t)-\int_t^T \alpha^{(i)}(t,s)ds+\frac{1}{2}\Bigl|\Bigl|
\int_t^T \sigma_c^{(i)}(t,s)ds\Bigr|\Bigr|^2\right\}dt \nn \\
&& -\left(\int_t^T \sigma_c^{(i)}(t,s)ds\right)\cdot dW_t^{\mbb{Q}^{(i)}}~.
\eea
From the definition of the zero coupon bond $D^{(i)}$ in Theorem~\ref{theorem1}, 
$\{D^{(i)}(t,T)/\beta_c^{(i)}(t), t\in[0,T]\}$ must be a $\mbb{Q}^{(i)}$-martingale. 
This requires $D^{(i)}$'s drift to be equal to $c^{(i)}$, which yields
\bea
\alpha^{(i)}(t,s)=\sigma_c^{(i)}(t,s)\cdot \left(\int_t^s \sigma_c^{(i)}(t,u)du\right)~.
\eea
This gives the desired result.
\end{proof}
\end{proposition}
\subsection{Dynamics of the forward LIBOR-OIS spread}
\label{sec-HJM-LOIS}
We denote the LIBOR of currency $(i)$ fixed at $T_{n-1}$ and maturing at $T_n$
as $L^{(i)}(T_{n-1},T_n)$.  Instead of modeling $L^{(i)}$ directly, we consider 
the dynamics of LIBOR-OIS spread, which is defined by
\bea
B^{(i)}(T_{n-1},T_n)=L^{(i)}(T_{n-1},T_n)-\frac{1}{\del^{(i)}_n}\left(\frac{1}{D^{(i)}(T_{n-1},T_n)}-1\right)
\eea
where $\del^{(i)}_n$ denotes the day-count fraction of $L^{(i)}$ for the period of $[T_{n-1},T_n]$.
It is clear that both $L^{(i)}(T_{n-1},T_n)$ and $B^{(i)}(T_{n-1},T_n)$ are
$\calf_{T_{n-1}}$ measurable.
\begin{definition}
The forward LIBOR-OIS spread for the period $[T_{n-1},T_n]$ is defined by
\bea
B^{(i)}(t;T_{n-1},T_n)&:=&\mbb{E}_t^{\mbb{T}^{(i)}_n}\left[B^{(i)}(T_{n-1},T_n)\right] \nn \\
&=&\mbb{E}_t^{\mbb{T}^{(i)}_n}\left[L^{(i)}(T_{n-1},T_n)\right]-\frac{1}{\del^{(i)}_n}\left(
\frac{D^{(i)}(t,T_{n-1})}{D^{(i)}(t,T_n)}-1\right)~.
\eea
\end{definition}
Here, the Radon-Nikodym density for the forward measure is given, as before, by 
\be
\left.\frac{d\mbb{T}^{(i)}_n}{d\mbb{Q}^{(i)}}\right|_t=\frac{D^{(i)}(t,T_n)}{\beta_c^{(i)}(t)D^{(i)}(0,T_n)}~.
\ee

Let us denote the volatility process for $B^{(i)}(\cdot;T_{n-1},T_n)$ by some 
appropriate adapted process $\sigma_B^{(i)}(\cdot;T_{n-1},T_n):\Omega\times \mbb{R}_{+}\rightarrow \mbb{R}^d$.
\begin{proposition}
Under the setup given in Sections 3.1 and 3.2, the no-arbitrage dynamics of 
the LIBOR-OIS spread $\{B^{(i)}(t;T_{n-1},T_n),t\in[0,T_{n-1}]\}$ for every $0\leq T_{n-1}<T_n$ is given by
\bea
\frac{dB^{(i)}(t;T_{n-1},T_n)}{B^{(i)}(t;T_{n-1},T_n)}=\sigma_B^{(i)}(t;T_{n-1},T_n)\cdot
\left(\int_t^{T_n}\sigma_c^{(i)}(t,s)ds\right)dt+\sigma_B^{(i)}(t;T_{n-1},T_n)\cdot dW_t^{\mbb{Q}^{(i)}}~\nn \\
\eea
with 
\be
B^{(i)}(0;T_{n-1},T_n)=\mbb{E}^{\mbb{T}^{(i)}_n}\Bigl[L^{(i)}(T_{n-1},T_n)\Bigr]-\frac{1}{\del_n^{(i)}}
\left(\frac{D^{(i)}(0,T_{n-1})}{D^{(i)}(0,T_n)}-1\right)~.
\ee
\begin{proof}

By construction, $B^{(i)}(\cdot;T_{n-1},T_n)$ is a martingale under the forward measure $\mbb{T}_n^{(i)}$.
Thus, in general, one can write its dynamics as
\be
dB^{(i)}(t;T_{n-1},T_n)/B^{(i)}(t;T_{n-1},T_n)=\sigma_B^{(i)}(t;T_{n-1},T_n)\cdot dW_t^{\mbb{T}_n^{(i)}}
\ee
where $W_t^{\mbb{T}_n^{(i)}}$ is a $d$-dimensional $\mbb{T}^{(i)}_n$-Brownian motion.
By Maruyama-Girsanov theorem, one can check that the relation
\bea
dW_t^{\mbb{T}_n^{(i)}}=dW_t^{\mbb{Q}^{(i)}}+\left(\int_t^{T_n}\sigma_c^{(i)}(t,s)ds\right)dt
\eea
holds, which gives the desired result.
\end{proof}
\end{proposition}
\subsection{Dynamics of the forward currency funding spread}
\label{sec-HJM-yij}
The remaining important ingredient for the term structure modeling is the 
dynamics of the currency funding spread, $y^{(i,j)}$. 
To understand the direct relationship between the observed cross currency basis 
and this currency funding spread, see \cite{ccc-orig}.
It has shown that the funding spread $y^{(i,j)}$ is the main cause of the currency basis
by analyzing the actual market data.
\begin{definition}
\label{def-fwdy}
The instantaneous forward funding spread at time $t$ with maturity $T\geq t$ 
of currency $(j)$ with respect to currency $(i)$ is defined by
\be
y^{(i,j)}(t,T):=-\frac{\part}{\part T}\ln Y^{(i,j)}(t,T)
\ee
where
\bea
\label{eq-defYij}
Y^{(i,j)}(t,T):=\mbb{E}_t^{\mbb{T}^{(i)}}\left[e^{-\int_t^T y^{(i,j)}(s)ds}\right]~.
\eea
\end{definition}

Let us denote the volatility process of the forward funding spread $y^{(i,j)}$
by $\sigma^{(i,j)}:\Omega\times \mbb{R}_{+}\times \mbb{R}_{+} \rightarrow \mbb{R}^d$
where $\{\sigma_y^{(i,j)}(t,s),t\in[0,s]\}$ is $\mbb{F}$-adapted for every $s\geq 0$.
\begin{proposition}\footnote{Notice that in Eq (6.31) and hence also in (6.39) of \cite{RiskBook}, the third component
$\sigma^{(i)}_c(t,s)\cdot \int_t^s \sigma_y^{(i,j)}(t,u)du$ of the drift term is missing, which should be 
corrected as in this proposition.}
Under the setup given in Sections 3.1 and 3.3, the no-arbitrage dynamics of the 
forward funding spread $\{y^{(i,j)}(t,s), t\in[0,s]\}$ for every $s\geq 0$ is given by
\bea
&&dy^{(i,j)}(t,s)=\left\{\sigma_y^{(i,j)}(t,s)\cdot \int_t^s \sigma_y^{(i,j)}(t,u)du+
\sigma_y^{(i,j)}(t,s)\cdot \int_t^s \sigma_c^{(i)}(t,u)du\right\}\nn \\
&&\qquad\qquad  \left. +\sigma_c^{(i)}(t,s)\cdot \int_t^s\sigma_y^{(i,j)}(t,u)du\right\}dt
+\sigma_y^{(i,j)}(t,s)\cdot dW_t^{\mbb{Q}^{(i)}} 
\eea
with  $y^{(i,j)}(t,t)=y^{(i,j)}(t)$~.
\begin{proof}
Define
\bea
\wt{Y}^{(i,j)}(t,T):=\mbb{E}^{\mbb{Q}^{(i)}}_t\left[e^{-\int_t^T (c^{(i)}(s)+y^{(i,j)}(s))ds}\right]
\label{def-yijtil}
\eea
and the corresponding forward rate $\{\wt{y}^{(i,j)}(t,T),t\in[0,T]\}$  as
\be
\wt{y}^{(i,j)}(t,T)=-\frac{\part}{\part T}\ln \wt{Y}^{(i,j)}(t,T)~.
\ee
Notice that
\bea
&&\wt{y}^{(i,j)}(t,T)=-\frac{\part}{\part T}\ln \wt{Y}^{(i,j)}(t,T)\nn \\
&&\quad =\frac{1}{\wt{Y}^{(i,j)}(t,T)}\mbb{E}^{\mbb{Q}^{(i)}}_t\left[
e^{-\int_t^T (c^{(i)}(s)+y^{(i,j)}(s))ds}(c^{(i)}(T)+y^{(i,j)}(T))\right]\nn\\
\eea
and hence using the fact that $\wt{Y}^{(i,j)}(t,t)=1$, one obtains by passing to the limit $T\downarrow t$,
\bea
\wt{y}^{(i,j)}(t,t)=c^{(i)}(t)+y^{(i,j)}(t)~.
\eea

Using the above result and the fact that
\bea
\left\{e^{-\int_0^t (c^{(i)}(s)+y^{(i,j)}(s))ds}\wt{Y}^{(i,j)}(t,T), ~t\in[0,T]\right\}
\eea
is a $\mbb{Q}^{(i)}$-martingale for every $T\geq 0$, one obtains, for $t\in[0,s]$ for every $s\geq 0$, that
\bea
&&d\wt{y}^{(i,j)}(t,s)=\wt{\sigma}^{(i,j)}(t,s)\cdot \Bigl(\int_t^s \wt{\sigma}^{(i,j)}(t,u)du\Bigr)dt
+\wt{\sigma}^{(i,j)}(t,s)\cdot dW_t^{\mbb{Q}^{(i)}} \\
&&\wt{y}^{(i,j)}(t,t)=c^{(i)}(t)+y^{(i,j)}(t)
\eea
from exactly the same arguments in Proposition~\ref{prop-cfwd}.
Here, $\wt{\sigma}^{(i,j)}:\Omega\times\mbb{R}_{+}\times\mbb{R}_{+}\rightarrow \mbb{R}^d$
corresponds to some volatility process, and $\{\wt{\sigma}^{(i,j)}(t,s),t\in[0,s]\}$
is $\mbb{F}$-adapted.

Now, let us decompose $\wt{y}^{(i,j)}$ into two parts as
\bea
\wt{y}^{(i,j)}(t,s)=c^{(i)}(t,s)+y^{(i,j)}(t,s)~.
\eea
From the definition of the forward collateral rate, one sees the second component satisfies
\be
\exp\Bigl(-\int_t^T y^{(i,j)}(t,s) ds\Bigr)=Y^{(i,j)}(T)
\ee
which is consistent with Definition~\ref{def-fwdy}.

One can write a general dynamics of the funding spread as
\bea
dy^{(i,j)}(t,s)=\mu^{(i,j)}(t,s)dt+\sigma^{(i,j)}_y(t,s)\cdot dW_t^{\mbb{Q}^{(i)}}~
\eea
using an appropriate adapted drift process $\{\mu^{(i,j)}(t,s), t\in[0,s]\}$.
Then, by construction,  one must have
\bea
\wt{\sigma}^{(i,j)}(t,s)=\sigma_c^{(i)}(t,s)+\sigma_y^{(i,j)}(t,s)~.
\eea
By taking the difference $dy^{(i,j)}(t,s)=d\wt{y}^{(i,j)}(t,s)-dc^{(i)}(t,s)$, one obtains
\bea
\mu^{(i,j)}(t,s)=\sigma_y^{(i,j)}(t,s)\cdot \int_t^s \bigl[\sigma_y^{(i,j)}(t,u)+
\sigma_c^{(i)}(t,u)\bigr]du+\sigma_c^{(i)}(t,s)\cdot \int_t^s\sigma_y^{(i,j)}(t,u)du
\eea
for every $s\geq 0$. This gives the desired result.
\end{proof}
\end{proposition}

\subsection{Dynamics of the spot foreign exchange rate}
\label{sec-HJM-fx}
Now, the last piece for the term structure modeling is the dynamics of the spot foreign exchange rate $f_x^{(i,j)}$.
Since we have the relation
\be
r^{(i)}_t-r^{(j)}_t=c^{(i)}_t-c^{(j)}_t+y^{(i,j)}_t
\ee
it is easy to see that the relevant dynamics is given by
\bea
df_x^{(i,j)}(t)/f_x^{(i,j)}(t)=\Bigl(c^{(i)}_t-c^{(j)}_t+y^{(i,j)}_t\Bigr)dt+\sigma_X^{(i,j)}(t)\cdot dW_t^{\mbb{Q}^{(i)}}~.
\eea
where $\sigma_X^{(i,j)}:\Omega\times \mbb{R}\rightarrow \mathbb{R}^d$ 
is an appropriate adapted process for the spot FX volatility.

The Radon-Nikodym density between two money-market measures with different currencies are given by
\bea
\left.\frac{d\mbb{Q}^{(j)}}{d\mbb{Q}^{(i)}}\right|_t&=&\frac{\beta^{(j)}(t) f_x^{(i,j)}(t)}{\beta^{(i)}(t)f_x^{(i,j)}(0)}=\frac{\beta_c^{(j)}(t)f_x^{(i,j)}(t)}{\beta_c^{(i)}(t)\beta_y^{(i,j)}(t)f_x^{(i,j)}(0)}
\label{RN-FX}
\eea
where $\beta_y^{(i,j)}(t)=e^{\int_0^t y^{(i,j)}_s ds}$~.  Since we have the relation
\bea
dW_t^{\mbb{Q}^{(j)}}=dW_t^{\mbb{Q}^{(i)}}-\sigma_X^{(i,j)}(t)dt~,
\eea
it is easy to change the currency measure. For example, one can check that the dynamics of the forward collateral rate
of currency $(j)$ becomes
\bea
dc^{(j)}(t,s)=\sigma_c^{(j)}(t,s)\cdot \left[\left(\int_t^s \sigma_c^{(j)}(t,u)du\right)-\sigma_X^{(i,j)}(t)\right]dt
+\sigma_c^{(j)}(t,s)\cdot dW_t^{\mbb{Q}^{(i)}}~
\eea
under the money-market measure of currency $(i)$. 

\subsection{Summary of the dynamics}
From Sections~\ref{sec-HJM-ois} to \ref{sec-HJM-fx}, we have derived the dynamics of 
all the relevant processes in the HJM framework.
For the convenience of readers, let us summarize the resultant system of stochastic differential equations (SDEs).
Here, we set currency $(i)$ as the base currency.
\\\\
{\bf{Rates for the base currency} }
\bea
&&dc^{(i)}(t,s)=\sigma_c^{(i)}(t,s)\cdot\left(\int_t^s \sigma_c^{(i)}(t,u)du\right)dt+\sigma_c^{(i)}(t,s)\cdot dW_t^{\mbb{Q}^{(i)}}\\
&&\frac{dB^{(i)}(t;T_{n-1},T_n)}{B^{(i)}(t;T_{n-1},T_n)}=\sigma_B^{(i)}(t;T_{n-1},T_n)\cdot
\left(\int_t^{T_n}\sigma_c^{(i)}(t,s)ds\right)dt+\sigma_B^{(i)}(t;T_{n-1},T_n)\cdot dW_t^{\mbb{Q}^{(i)}} \nn \\
\eea
{\bf{Funding spreads with respect to the base currency}}
\bea
&&dy^{(i,j)}(t,s)=\left\{\sigma_y^{(i,j)}(t,s)\cdot\Bigl( \int_t^s \bigl[\sigma_y^{(i,j)}(t,u)+
 \sigma_c^{(i)}(t,u)\bigr]du\Bigr)\right.\nn \\
&&\qquad\qquad  \left. +\sigma_c^{(i)}(t,s)\cdot \int_t^s\sigma_y^{(i,j)}(t,u)du\right\}dt
+\sigma_y^{(i,j)}(t,s)\cdot dW_t^{\mbb{Q}^{(i)}} 
\eea
{\bf{Foreign exchange rate}}
\bea
df_x^{(i,j)}(t)/f_x^{(i,j)}(t)=\Bigl(c^{(i)}_t-c^{(j)}_t+y^{(i,j)}_t\Bigr)dt+\sigma_X^{(i,j)}(t)\cdot dW_t^{\mbb{Q}^{(i)}}
\eea
{\bf{Rates for a foreign currency}}
\bea
&&dc^{(j)}(t,s)=\sigma_c^{(j)}(t,s)\cdot \left[\left(\int_t^s \sigma_c^{(j)}(t,u)du\right)-\sigma_X^{(i,j)}(t)\right]dt
+\sigma_c^{(j)}(t,s)\cdot dW_t^{\mbb{Q}^{(i)}} \\
&&\frac{dB^{(j)}(t;T_{n-1},T_n)}{B^{(j)}(t;T_{n-1},T_n)}=\sigma_B^{(j)}(t;T_{n-1},T_n)\cdot
\left[\left(\int_t^{T_n}\sigma_c^{(i)}(t,s)ds\right)-\sigma_X^{(i,j)}(t)\right]dt\nn\\
&&\hspace{37mm} +\sigma_B^{(i)}(t;T_{n-1},T_n)\cdot dW_t^{\mbb{Q}^{(i)}}
\eea
{\bf{Funding spreads with respect to a foreign currency}}
\bea
&&dy^{(j,k)}(t,s)=\left\{\sigma_y^{(j,k)}(t,s)\cdot\Bigl( \int_t^s \bigl[\sigma_y^{(j,k)}(t,u)+
 \sigma_c^{(j)}(t,u)\bigr]du-\sigma_X^{(i,j)}(t)\Bigr)\right.\nn \\
&&\qquad\qquad  \left. +\sigma_c^{(j)}(t,s)\cdot \int_t^s\sigma_y^{(j,k)}(t,u)du\right\}dt
+\sigma_y^{(j,k)}(t,s)\cdot dW_t^{\mbb{Q}^{(i)}} 
\eea

\section{Some remarks on the formulation in \cite{ccc-old,ccc-orig}}
In our previous works, we have defined the instantaneous forward rate
of the funding spread as
\bea
e^{\int_t^T y^{(i,j)}(t,s)ds}=\mbb{E}_t^{\mbb{Q}^{(i)}}\Bigl[e^{-\int_t^T y^{(i,j)}(s)ds}\Bigr]
\eea
or equivalently,
\bea
y^{(i,j)}(t,T)=-\frac{\part}{\part T} \ln \mbb{E}_t^{\mbb{Q}^{(i)}}\Bigl[e^{-\int_t^T y^{(i,j)}(s)ds}\Bigr]~.
\eea
Notice the difference in the associated measure in Definition~\ref{def-fwdy}.
In the above definition, the dynamics of the funding spread is actually simpler:
\bea
dy^{(i,j)}(t,s)=\sigma_y^{(i,j)}(t,s)\cdot \left(\int_t^s \sigma_y^{(i,j)}(t,u)du\right)dt+
\sigma_y^{(i,j)}(t,s)\cdot dW_t^{\mbb{Q}^{(i)}}.
\eea

However, the above formulation makes the calibration procedures rather complicated 
in the presence of non-zero correlation between the collateral rate and the funding spread.
Suppose that there is a currency-$(j)$ collateralized zero coupon bond denominated by 
currency $(i)$. This is an important building block for the pricing of
the collateralized contracts with a foreign currency.
The present value of this bond is 
\bea
\mbb{E}_t^{\mbb{Q}^{(i)}}\Bigl[e^{-\int_t^T (c^{(i)}(s)+y^{(i,j)}(s))ds}\Bigr]~.
\eea
Due to the non-zero correlation, the above quantity depends on the 
dynamics of both of the forward rates,
and is impossible to separate the two effects.
On the other hand , in the formulation proposed in Section~\ref{sec-HJM-yij},
the above quantity is given by $D^{(i)}(t,T)Y^{(i,j)}(t,T)$ without 
any independence assumption (See Eqs.(\ref{f_pricing_formula2}) and (\ref{eq-defYij}).). In addition, if the market quotes are available for this type of products, 
then one can easily get the clean information for $Y^{(i,j)}$
which directly gives the initial condition for the funding spread by differentiation.

\section{Collateralized FX}
\label{sec-fx}

\subsection{Collateralized FX forward}
\label{sec-fxforward}
Let first consider a collateralized FX forward contract between currency $(i)$ and $(j)$,
in which a unit amount of currency $(j)$ is going to be exchanged by $K$ units amount of 
currency $(i)$ at the maturity $T$. The amount of $K$ is fixed at $t$, the trade inception time.
Assume the contract is fully collateralized by currency $(k)$.
Here, the amount of $K$ that makes this exchange have the present value $0$ at time $t$ is 
called the forward FX rate. 

The break-even condition for the amount of $K$ is given by
\bea
K\mbb{E}_t^{\mbb{Q}^{(i)}}\left[e^{-\int_t^T (c_s^{(i)}+y_s^{(i,k)})ds}\right]=f_x^{(i,j)}(t)
\mbb{E}_t^{\mbb{Q}^{(j)}}\left[e^{-\int_t^T (c_s^{(j)}+y_s^{(j,k)})ds}\right]~.
\eea
Here, the left-hand side represents the value of $K$ units amount of currency $(i)$ paid at $T$
with collateralization in currency $(k)$. In the right-hand side the value of a unit amount of 
currency $(j)$ is represented in terms of currency $(i)$ by multiplying the spot FX rate.
By solving the equation for $K$, we obtain the forward FX as (See definition (\ref{def-yijtil}).)
\bea
\label{eq-fc-fx}
f_x^{(i,j)}(t;T,(k))=f_x^{(i,j)}(t)\frac{\wt{Y}^{(j,k)}(t,T)}{\wt{Y}^{(i,k)}(t,T)}
\eea
where the last argument $(k)$ denotes the currency used as the collateral.
In general, the currency triangle relation among forward FXs only holds with the 
common collateral currency:
\be
f_x^{(i,j)}(t,T;(k))\times f_x^{(j,l)}(t,T;(k))=f_x^{(i,l)}(t,T;(k))~.
\ee

Suppose the same contract is made with a collateral currency either $(i)$ or $(j)$,
which seems more natural in the market.
In this case, we have one-to-one relation between the forward FX value and 
the forward funding spread. For example, if currency $(i)$ is used as
the collateral, we have
\bea
\label{eq-fwdfx}
f_x^{(i,j)}(t;T,(i))=f_x^{(i,j)}(t)\frac{D^{(j)}(t,T)}{D^{(i)}(t,T)} Y^{(j,i)}(t,T).
\eea
Therefore, if we can observe the forward FX with various maturities in the market, we can bootstrap the 
forward curve of $\{y^{(j,i)}(t,\cdot)\}$ since we already know $D^{(i)}$ and $D^{(j)}$
from each OIS market. When collateralization is done by currency $(j)$,
we can extract $\{y^{(i,j)}(t,\cdot)\}$ by the same arguments.

When the collateral currency is $(i)$, one can easily show that
\bea
f_x^{(i,j)}(t)\mbb{E}_t^{\mbb{Q}^{(j)}}\Bigl[e^{-\int_t^T(c_s^{(j)}+y_s^{(j,i)})ds}\Bigr]=\mbb{E}_t^{\mbb{Q}^{(i)}}
\Bigl[e^{-\int_t^T c^{(i)}_s ds}f_x^{(i,j)}(T)\Bigr]
\eea
by using the Radon-Nikodym density (\ref{RN-FX}).
Thus, we see
\bea
f_x^{(i,j)}(t;T,(i))=\mbb{E}_t^{\mbb{T}^{(i)}}\Bigl[f_x^{(i,j)}(T)\Bigr]
\eea
from Eq.~(\ref{eq-fwdfx}). Hence, the forward FX collateralized by its domestic currency $(i)$ is a martingale
under the associated forward measure $\mbb{T}^{(i)}$. In more general situation, we have the expression
\bea
f_x^{(i,j)}(t;T,(k))=\mbb{E}_t^{\mbb{T}^{(i,k)}}\Bigl[f_x^{(i,j)}(T)\Bigr]
\label{fxforward_f_collateral}
\eea
where the new forward measure is defined by 
\bea
\left.\frac{d\mbb{T}^{(i,k)}}{d\mbb{Q}^{(i)}}\right|_t=\frac{\wt{Y}^{(i,k)}(t,T)}{\beta_c^{(i)}(t)\beta_y^{(i,k)}(t)
\wt{Y}^{(i,k)}(0,T)}~,
\label{foreign_forward}
\eea
and $\{f_x^{(i,j)}(t;T,(k)), t\in[0,T]\}$ is a $\mbb{T}^{(i,k)}$-martingale.
One can see, $\wt{Y}^{(i,k)}(t,T)$, the price of a zero-coupon bond of currency $(i)$ collateralized by currency $(k)$ 
is associated as a numeraire for the forward measure $\mbb{T}^{(i,k)}$.

\subsection{FX option}
\label{sec-fxoption}
Just for completeness, we give a brief remark on collateralized FX European options.
Let us consider a $T$-maturing European call option on $f_x^{(i,j)}$ with strike $K$
collateralized by a currency $(k)$.
The present value of the option is
\be
PV_t=\mbb{E}_t^{\mbb{Q}^{(i)}}\left[e^{-\int_t^T (c^{(i)}_s+y^{(i,k)}_s)ds}\Bigl(f_x^{(i,j)}(T)-K\Bigr)^+\right]~,
\ee
where $X^+$ denotes $\max(X,0)$.
Since we can rewrite it as
\be
PV_t=D^{(i)}(t,T)Y^{(i,k)}(t,T)\mbb{E}_t^{\mbb{T}^{(i,k)}}\left[\Bigl(f_x^{(i,j)}(T;T,(k))-K\Bigr)^+\right]~
\ee
it is enough to know the dynamics of $f_x(t;T,(k))$ under the measure $\mbb{T}^{(i,k)}$.
By applying It\^o formula to (\ref{eq-fc-fx}), we obtain
\bea
df_x^{(i,j)}(t;T,(k))/f_x^{(i,j)}(t;T,(k))=\left\{
\sigma_X^{(i,j)}(t)+\Gamma^{(i,k)}(t,T)-\Gamma^{(j,k)}(t,T)\right\}\cdot dW_t^{\mbb{T}^{(i,k)}}
\eea
where $W^{\mbb{T}^{(i,k)}}$ is a $d$-dimensional $\mbb{T}^{(i,k)}$-Brownian motion. 
Here, for $m=\{i,j\}$, we have defined
\bea
\Gamma^{(m,k)}(t,T)=\int_t^T \Bigl(\sigma_c^{(m)}(t,s)+\sigma_y^{(m,k)}(t,s)\Bigr)ds~.
\eea
Thus, when both of the collateral and funding spread volatilities are
deterministic (hence follow Gaussian HJM), one can use Black's formula.
Although there is no closed-form solution except this special case,
asymptotic expansion technique can be applied to derive an analytical 
approximate solution. See, for example, Takahashi (2015)~\cite{T} and references therein.

\section{Implementation with a fixed tenor structure}
\label{sec-disc-HJM}
Due to the infinite dimensionality, a direct numerical implementation of the proposed HJM framework 
is impossible.  An obvious solution is to reduce its dimension by discretization following 
the idea of BGM (or LIBOR market) model~\cite{BGM}. 
One can see, for example, Mercurio~\cite{Mercurio}~(2009) as an early attempt.
In recent years, many researchers have followed  the same route
or adopted an Affine model for tractability.
However, a complete picture including multiple currencies
and currency funding spreads, which are necessary to explain the cross currency basis, 
have only been  found in continuous HJM framework given in our previous works \cite{dynamic_basis, ccc-old, ccc-orig}.

In the remainder of the paper, we shall give a 
directly implementable discretization for the complete HJM framework 
including multiple currencies and the associated funding spreads.
Due to the limited capacity for numerical simulation, using 
a daily span for the overnight rates looks difficult.
When using a more sparse time partition, considering a daily compounding nature of 
the OIS, it seems more useful to define a continuously compounded forward rate
with a finite accrual period. It also looks suitable for the modeling of the funding spread
since it is the dividend yield being paid everyday to the collateral holder.
In addition, it possesses a much stronger similarity to the HJM framework than 
a traditional BGM-like discretization. Thanks to this similarity, one can use a continuous 
HJM framework,   which allows  more concise treatments, for the  analytical study of the implemented model
with a fixed tenor structure.
\subsection{Setup}
\label{sec-setup}
We work in a filtered probability space $(\Omega,\calf,\mbb{F},\mbb{P})$ satisfying the usual conditions,
where $\mbb{F}$ is assumed to be an augmented filtration generated by the $d$-dimensional Brownian motion $W$.
It is assumed to support every stochastic variable to appear in the model.
We first introduce a partition of the time interval called a {\it tenor structure} as
\bea
0=T_0<T_1<T_2<\cdots <T_{N^*}~
\eea
and $\del_i:=T_{i+1}-T_{i}$, where $T_{N^*}$ is some fixed time horizon.
We also define 
\bea
q(t):=\min\Bigl\{n; T_n\geq t\Bigr\}
\eea
and hence we always have $T_{q(t)-1}<t \leq T_{q(t)}$.
We assume, for each currency,  that the traditional risk-free zero coupon bonds
and the collateralized zero coupon bonds are continuously tradable at any time $t\in[0,T_{N^*}]$.
Their price processes are assumed to be strictly positive and denoted by
\bea
P^{(i)}(t,T_{q(t)}),\cdots, P^{(i)}(t,T_{N^*})
\eea
and
\bea
D^{(i)}(t,T_{q(t)}),\cdots, D^{(i)}(t,T_{N^*})
\eea
for each currency.

Following the idea of Schl\"ogl~\cite{FXLMM}, for every currency $(i)$, we assume that 
the prices of the shortest bonds $P^{(i)}(t,T_{q(t)})$ $D^{(i)}(t,T_{q(t)})$
are $\calf_{T_{q(t)-1}}$-measurable and 
absolutely continuous with respect to the Lebesgue measure. 
Then, we can define the associated short rate processes $(r^{(i)}, c^{(i)})$  
in such a way that
\bea
\label{eq-r}
&&\exp\Bigl(-\int_t^{T_{q(t)}}r^{(i)}(s)ds\Bigr)=P^{(i)}(t,T_{q(t)}) \\
&&\exp\Bigl(-\int_t^{T_{q(t)}}c^{(i)}(s)ds\Bigr)=D^{(i)}(t,T_{q(t)})
\eea
for every $t\in[0,T_{N^*}]$, and $(r^{(i)}, c^{(i)})$ are $\calf_{T_{q(t)-1}}$-measurable
processes.
We denote
\be
y^{(i)}(t):=r^{(i)}(t)-c^{(i)}(t), \quad t\in[0,T_{N^*}]
\ee
which corresponds to the dividend yield associated to the currency-$(i)$ collateralized
products.

Let us first define a discrete bank account for a given currency $(i)$:
\bea
B^{(i)}(t):=\exp\Bigl(\sum_{m=0}^{q(t)-1} \del_m f^{(i)}_m(T_m)\Bigr)P^{(i)}(t,T_{q(t)})~.
\eea
Here,
\be
f_m^{(i)}(t):=-\frac{1}{\del_m}\ln \left(\frac{P^{(i)}(t,T_{m+1})}{P^{(i)}(t,T_m)}\right), t\in[0,T_m]
\ee 
is a forward rate at time $t$ for the period $[T_m,T_{m+1}]$ with a continuous 
compounding~\footnote{If we adopt a simple rate convention, we obtain a traditional BGM model.}.
It is easy to show that the market is arbitrage free if there exists a measure $\mbb{Q}_B^{(i)}$
equivalent to $\mbb{P}$ such that every (non-dividend paying) asset price denominated in 
currency $(i)$ discounted by $B^{(i)}$ is a 
$\mbb{Q}_B^{(i)}$-martingale.
\subsubsection*{Remark}
Suppose we create the continuous money market account $\beta^{(i)}(t)=\exp\Bigl(\int_0^t r^{(i)}(s)ds\Bigr)$
by using the rate process defined in (\ref{eq-r}).
Then the new measure $\mbb{Q}^{(i)}$ associated with this $\beta^{(i)}$ as the numeraire,
is actually {\it equal} to $\mbb{Q}_B^{(i)}$ since both numeraires have finite variation.
\\ 
 
Although the above numeraire asset is conceptually useful, it remains completely implicit in 
a collateralized market. In the presence of collateralization, the fundamental assets
are  dividend yielding assets.
Let us define a discrete collateral account by
\bea
C^{(i)}(t):=\exp\Bigl(\sum_{m=0}^{q(t)-1}\del_m c_m^{(i)}(T_m)\Bigr)D^{(i)}(t,T_{q(t)})~,
\label{eq-disc-cacount}
\eea
where
\be
c_m^{(i)}(t):=-\frac{1}{\del_m}\ln \left(\frac{D^{(i)}(t,T_{m+1})}{D^{(i)}(t,T_m)}\right), t\in[0,T_m]~
\ee
is the corresponding collateral forward rate.

Now, let us choose a certain base currency and omit its specification $(i)$.
Consider the market that contains $m$ non-dividend paying assets
and $n$ $y$-yielding collateralized assets. Let us denote their price processes
by $\{S_i(t),t\in[0,T_{N^*}]\}$ and $\{U_j(t), t\in[0,T_{N^*}]\}$
for $i\in\{1,\cdots,m\}$ and $j\in\{1,\cdots,n\}$.
\begin{proposition}
\label{no-arb}
The market is arbitrage free if $\{S_i(t)/B(t),t\in[0,T_{N^*}]\}_{1\leq i\leq m}$ and
$\{U_j(t)/C(t),t\in[0,T_{N^*}]\}_{1\leq j\leq n}$ are all $\mbb{Q}_B$-martingales.
\begin{proof}
Let us denote the self-financing trading strategy by $\{(\pi_i(t),\eta_i(t)),t\in[0,T_{N^*}]\}$, 
$i\in\{1,\cdots,m\}$, $j\in\{1,\cdots,n\}$.
The associated wealth process $\{V(t),t\in[0,T_{N^*}]\}$ is given by
\bea
&&dV(t)=\pi(t) \cdot dS(t)+\eta(t)\cdot dU(t)+y(t)(\eta(t)\cdot U(t))dt\nn \\
&&\quad+\bigl[V_t-\pi(t)\cdot S(t)-\eta(t)\cdot U(t)\bigr]
\frac{dB(t)}{B(t)}~.
\eea
Note that the investments into the collateralized zero coupon bonds require
continuous reinvestments of their dividends, which eventually leads to the same expression.

For the absence of arbitrage, it is enough to show that $\{V(t)/B(t), t\in[0,T_{N^*}]\}$ is a $\mbb{Q}_B$-martingale.
It is easy to see
\bea
&&d\left(\frac{V(t)}{B(t)}\right)=\frac{dV(t)}{B(t)}-\frac{V(t)}{B(t)^2}dB(t) \nn \\
&&=\pi_t\cdot \Bigl(\frac{dS(t)}{B(t)}-\frac{S(t)}{B(t)^2}dB(t)\Bigr)
+\frac{C(t)}{B(t)}\eta(t)\cdot \Bigl(\frac{dU(t)}{C(t)}-\frac{U(t)}{C(t)^2}dC(t)\Bigr)\nn \\
&&+\frac{\eta(t)\cdot U(t)}{B(t)}\Bigl(\frac{dC(t)}{C(t)}+y(t)dt-\frac{dB(t)}{B(t)}\Bigr)~.
\eea
Since the last term vanishes, one obtains
\bea
d\left(\frac{V(t)}{B(t)}\right)=\pi(t)\cdot d\Bigl(\frac{S(t)}{B(t)}\Bigr)+\frac{C(t)}{B(t)}\eta(t)
\cdot d\Bigl(\frac{U(t)}{C(t)}\Bigr)
\eea
which proves the desired result.
\end{proof}
\end{proposition}
The above result implies that one can guarantee the absence of arbitrage
in a market collateralized by the domestic currency $(i)$ by 
imposing that every $C^{(i)}$-discounted price process is a $\mbb{Q}_B^{(i)}$-martingale.
The arguments is easily extendable to the situation where there exist
collateralized products by a foreign currency $(j)$.
In this market, one can additionally trade the set of domestic zero-coupon bonds
with the same tenor structure but collateralized by the foreign currency $(j)$:
\be
\wt{Y}^{(i,j)}(t,T_{q(t)}), \cdots, \wt{Y}^{(i,j)}(t,T_{N^*})~,
\ee
which are assumed to have a common dividend yield $y^{(j)}=r^{(j)}-c^{(j)}$. 
The shortest bond satisfies 
\bea
\wt{Y}^{(i,j)}(t,T_{q(t)})&=&\exp\Bigl(-\int_t^{T_{q(t)}}(r^{(i)}(s)-y^{(j)}(s))ds\Bigr)\nn \\
&=&\exp\Bigl(-\int_t^{T_{q(t)}}(c^{(i)}(s)+y^{(i,j)}(s))ds\Bigr)
\eea
where $y^{(i,j)}=y^{(i)}-y^{(j)}$ and are $\calf_{T_{q(t)-1}}$-measurable. 
Let us define a new collateralized account by the currency $(j)$ as
\bea
\label{disc-bank-yij}
C^{(i,j)}(t)=\exp\Bigl(\sum_{m=0}^{q(t)-1}\del_m (c_m^{(i)}(T_m)+y_m^{(i,j)}(T_m))\Bigr)\wt{Y}^{(i,j)}(t,T_{q(t)})~.
\eea
with
\bea
&&y_m^{(i,j)}(t)=-\frac{1}{\del_m}\ln \left(\frac{Y^{(i,j)}(t,T_{m+1})}{Y^{(i,j)}(t,T_m)}\right) \\
&&Y^{(i,j)}(t,T_m):=\wt{Y}^{(i,j)}(t,T_m)/D^{(i)}(t,T_m)~.
\eea
It is clear that $Y^{(i,j)}(t,T_{q(t)})$ is $\calf_{T_{q(t)-1}}$-measurable.
By exactly the same arguments in Proposition~\ref{no-arb}, 
the absence of arbitrage is guaranteed by requiring that every
price process of $(j)$-collateralized $(i)$-denominated asset divided by $C^{(i,j)}$
is a $\mbb{Q}_B^{(i)}$-martingale.

\subsection{Dynamics of discretized collateral rates}
We want to derive the dynamics of $\{c_m^{(i)}\}_{m\in\{0,\cdots,N^*-1\}}$ under $\mbb{Q}^{(i)}_B$.
Its dynamics in our Brownian setup can generally be represented by
\bea
dc_m^{(i)}(t)=\alpha^{(i)}_m(t)dt+\sigma^{(i)}_m(t)\cdot dW_t^{\mbb{Q}^{(i)}_B}
\eea
for $m\in\{0,\cdots,N^*-1\}$. Here, $\alpha_m^{(i)}:\Omega\times [0,T_m] \rightarrow \mbb{R}$
and $\sigma^{(i)}_m:\Omega\times [0,T_m]\rightarrow \mbb{R}$ 
are $\mbb{F}$-adapted processes. We define $c_m^{(i)}(t)=c_m^{(i)}(T_m)$ for $t\geq T_m$.

\begin{proposition}
\label{prop-cdisc}
The no-arbitrage dynamics of the collateral forward rate is given by
\bea
dc_n^{(i)}(t)=\left\{\sigma_n^{(i)}(t)\cdot \left(\sum_{m=q(t)}^{n-1}\del_m \sigma_m^{(i)}(t)\right)+
\frac{1}{2}\del_n |\sigma_n^{(i)}(t)|^2 \right\}dt+\sigma_n^{(i)}(t)\cdot dW_t^{\mbb{Q}_B^{(i)}}~
\eea
for $t\in[0,T_n]$ with every $n\in\{0,\cdots,N^*-1\}$.
\begin{proof}

For the absence of arbitrage, 
\bea
\frac{D^{(i)}(t,T_n)}{C^{(i)}(t)}, \quad t\in[0,T_n]
\label{eq-doverc}
\eea
must be a $\mbb{Q}^{(i)}_B$-martingale for every $n\in\{0,\cdots,N^*\}$.
It is not difficult to confirm that the ratio (\ref{eq-doverc})
is continuous at every $T_i, i\in\{0,\cdots,n\}$
by using the expression 
\bea
&&D^{(i)}(t,T_n)=D^{(i)}(t,T_{q(t)})\prod_{m=q(t)}^{n-1}\frac{D^{(i)}(t,T_{m+1})}{D^{(i)}(t,T_m)}\nn \\
&&\qquad =D^{(i)}(t,T_{q(t)})\exp\Bigl(-\sum_{m=q(t)}^{n-1}\del_m c_m^{(i)}(t)\Bigr),
\eea
and (\ref{eq-disc-cacount}).
Thus, it is enough to require 
\bea
X_n^{(i)}(t):=\exp\Bigl(-\sum_{m=q(t)}^{n-1}\del_m c_m^{(i)}(t)\Bigr)
\eea
to be a $\mbb{Q}^{(i)}_B$-martingale within each interval $(T_{q(t)-1},T_{q(t)})$ for every $n$.

An application of It\^o-formula in a given interval yields that
\bea
&&dX_n^{(i)}(t)=X_n^{(i)}(t)\left\{ -\sum_{m=q(t)}^{n-1}\del_m dc_m^{(i)}(t)+
\frac{1}{2}\sum_{m,m^\prime=q(t)}^{n-1}\del_m\del_{m^\prime}d\langle c_m^{(i)},c_{m^\prime}^{(i)}\rangle_t\right\}\nn \\
&&\qquad =X_n^{(i)}(t)\left\{-\sum_{m=q(t)}^{n-1}\del_m \alpha_m^{(i)}(t)+\frac{1}{2}
\Bigl|\sum_{m=q(t)}^{n-1}\del_m \sigma_m^{(i)}(t)\Bigr|^2\right\}dt\nn \\
&&\qquad -X_n^{(i)}(t)\sum_{m=q(t)}^{n-1}\del_m \sigma_m^{(i)}(t)\cdot dW_t^{\mbb{Q}_B^{(i)}}~.
\eea
This implies
\bea
\label{eq-noarb}
\sum_{m=q(t)}^{n-1}\del_m \alpha_m^{(i)}(t)=\frac{1}{2}\Bigl|\sum_{m=q(t)}^{n-1}\del_m \sigma_m^{(i)}(t)\Bigr|^2~.
\eea
Since this relation must hold for every $n\in\{0,\cdots,N^{*}-1\}$, one obtains,
by taking the difference, that
\bea
\del_n \alpha_n^{(i)}(t)&=&\frac{1}{2}\Bigl|\sum_{m=q(t)}^{n}\del_m \sigma_m^{(i)}(t)\Bigr|^2
-\frac{1}{2}\Bigl|\sum_{m=q(t)}^{n-1}\del_m \sigma_m^{(i)}(t)\Bigr|^2\nn \\
&=&\frac{1}{2}\del_n^2 |\sigma_n^{(i)}(t)|^2+\del_n \sigma_n^{(i)}(t)\cdot 
\Bigl(\sum_{m=q(t)}^{n-1}\del_m \sigma_m^{(i)}(t)\Bigr)~
\eea
and hence
\bea
\alpha_n^{(i)}(t)=\frac{1}{2}\del_n |\sigma_n^{(i)}(t)|^2+ \sigma_n^{(i)}(t)\cdot 
\Bigl(\sum_{m=q(t)}^{n-1}\del_m \sigma_m^{(i)}(t)\Bigr)~.
\eea
Once we fix $\alpha_n^{(i)}$ as above, one can recursively verify that
the relation (\ref{eq-noarb}) is satisfied for every $n$.
This proves the proposition.
\end{proof}
\end{proposition}

\subsection{Dynamics of collateralized LIBOR-OIS spreads}
The LIBOR-OIS spread $B^{(i)}(T_{n-1},T_n)$ itself is merely an index and not a tradable asset.
However, a collateralized forward contract is of course tradable and hence  
\bea
\mbb{E}_t^{\mbb{Q}_B^{(i)}}\left[e^{-\int_t^{T_n} c^{(i)}(s)ds}B^{(i)}(T_{n-1},T_n)\right]=D^{(i)}(t,T_n)B^{(i)}(t;T_{n-1},T_n),
~t\in [0,T_{n-1}]
\eea 
is a price process of a ``dividend $y^{(i)}$" yielding asset.

Let us write the general dynamics as
\bea
dB^{(i)}(t;T_{n-1},T_n):=B^{(i)}(t;T_{n-1},T_n)\left(b_n^{(i)}(t)dt+\sigma_{B,n}^{(i)}(t)\cdot dW_t^{\mbb{Q}_B^{(i)}}\right)~,
\eea
with some appropriate $\mbb{F}$-adapted processes $b_n^{(i)}:\Omega\times [0,T_{n-1}]\rightarrow \mbb{R}$ and $\sigma_{B,n}^{(i)}:\Omega\times[0,T_{n-1}]
\rightarrow \mbb{R}^d$.
\begin{proposition}
\label{prop-lois}
The no-arbitrage dynamics of the LIBOR-OIS forward spread is given by
\bea
dB^{(i)}(t;T_{n-1},T_n)=B^{(i)}(t;T_{n-1},T_n)\left(
\sigma_{B,n}^{(i)}(t)\cdot\Bigl(\sum_{m=q(t)}^{n-1}\del_m \sigma_m^{(i)}(t)\Bigr)dt+\sigma_{B,n}^{(i)}(t)
\cdot dW_t^{\mbb{Q}_B^{(i)}}\right)~\nn \\
\eea
for $t\in[0,T_{n-1}]$ with every $n\in\{0,\cdots,N^{*}\}$~.
\begin{proof}
For the absence of arbitrage, 
\bea
\frac{D^{(i)}(t,T_n)B^{(i)}(t;T_{n-1},T_n)}{C^{(i)}(t)}, \quad t\in[0,T_{n-1}]
\label{eq-dbratio}
\eea
must be a $\mbb{Q}^{(i)}_B$-arbitrage. This is equivalent to require the process
\bea
X_{B,n}^{(i)}(t):=\exp\Bigl(-\sum_{m=q(t)}^{n-1}\del_m c_m^{(i)}(t)\Bigr)B^{(i)}(t;T_{n-1},T_n)
\eea
to be a $\mbb{Q}^{(i)}_B$-martingale within an each interval $(T_{q(t)-1},T_{q(t)})$,
since (\ref{eq-dbratio}) is continuous at every node $T_i,i\in\{0,\cdots,n-1\}$.

An application of It\^o-formula in a given interval
and the result of Proposition~\ref{prop-cdisc}  yield
\bea
dX_{B,n}^{(i)}(t)&=&X_{B,n}^{(i)}(t)
\left(b_n^{(i)}(t)-\sigma_{B,n}^{(i)}(t)\cdot\Bigl(\sum_{m=q(t)}^{n-1}\del_m \sigma_m^{(i)}(t)\Bigr)\right)dt\nn \\
&&+X_{B,n}^{(i)}(t)\Bigl(\sigma_{B,n}^{(i)}(t)-\sum_{m=q(t)}^{n-1}\del_m \sigma_m^{(i)}(t)\Bigr)\cdot dW_t^{\mbb{Q}_B^{(i)}}~.
\eea
This implies 
\be
b_n^{(i)}(t)=\sigma_{B,n}^{(i)}(t)\cdot\Bigl(\sum_{m=q(t)}^{n-1}\del_m \sigma_m^{(i)}(t)\Bigr),
\ee
which proves the proposition~\footnote{
The same result can be obtained from the measure-change technique and 
the fact that $\{B^{(i)}(t;T_{n-1},T_n),t\in[0,T_{n-1}]\}$
is a martingale under the forward measure $\mbb{T}^{(i)}_n$.}.
\end{proof}
\end{proposition}

\subsection{Dynamics of currency funding spreads}
As in the previous section, the funding spread itself is not directly tradable
but a zero coupon bond collateralized by a foreign currency is a tradable asset.
Hence
\bea
\wt{Y}^{(i,j)}(t,T_n)&=&\mbb{E}^{\mbb{Q}_B^{(i)}}\left[e^{-\int_t^{T_n}(c^{(i)}(s)+y^{(i,j)}(s))ds}
\right]\nn \\
&&=D^{(i)}(t,T_n)Y^{(i,j)}(t,T_n), \quad t\in[0,T_n]
\eea
is a price process of a tradable asset. An important fact is that 
this asset has a stream of dividend yield
\be
y^{(j)}=y^{(i)}-y^{(i,j)}~.
\ee
According to the discussion in Section~\ref{sec-setup},
we have to discount it by $C^{(i,j)}(t)$ of (\ref{disc-bank-yij}) 
to impose a martingale condition.

Let us write the dynamics of $y_m^{(i,j)}$ as follows:
\bea
dy_m^{(i,j)}(t)=\alpha_m^{(i,j)}(t)dt+\sigma_{y,m}^{(i,j)}(t)\cdot dW_t^{\mbb{Q}_B^{(i)}}
\eea
where $\alpha_m^{(i,j)}:\Omega\times [0,T_{m}]\rightarrow \mbb{R}$
and $\sigma_m^{(i,j)}:\Omega\times[0,T_m]\rightarrow \mbb{R}^d$ are $\mbb{F}$-adapted processes.

\begin{proposition}
The no-arbitrage dynamics of the forward funding spread is given by
\bea
&&dy_n^{(i,j)}(t)=\left\{\frac{\bigl.}{\bigr.} \right. \sigma_{y,n}^{(i,j)}(t)\cdot \Bigl(\sum_{m=q(t)}^{n-1}\del_m 
[\sigma_{y,m}^{(i,j)}(t)+\sigma_m^{(i)}(t)]\Bigr)+\sigma_n^{(i)}(t)\cdot \Bigl(\sum_{m=q(t)}^{n-1}\del_m \sigma_{y,m}^{(i,j)}(t)\Bigr)\nn \\
&&\qquad\left. 
+\frac{1}{2}\del_n |\sigma_{y,n}^{(i,j)}(t)|^2+\del_n (\sigma_{y,n}^{(i,j)}(t)\cdot \sigma_n^{(i)}(t))\frac{\bigl.}{\bigr.}\right\}dt+\sigma_{y,n}^{(i,j)}(t)\cdot dW_t^{\mbb{Q}_B^{(i)}}~
\eea
for $t\in[0,T_n]$ with every $n\in\{0,\cdots,N^{*}-1\}$.
\begin{proof}
The absence of arbitrage requires
\bea
\frac{\wt{Y}^{(i,j)}(t,T_n)}{C^{(i,j)}(t)},\quad t\in[0,T_n]
\label{eq-yijcijratio}
\eea
to be a $\mbb{Q}_B^{(i)}$-martingale for every $n\in\{0,\cdots,N^*\}$.
As before, one can check that the ratio (\ref{eq-yijcijratio}) is 
continuous at every $T_i,i\in\{0,\cdots,n\}$.
Thus, this condition is equivalent to require,  for every $n\in\{0,\cdots,N^*\}$,
\bea
X_n^{(i,j)}(t):=\exp\left(-\sum_{m=q(t)}^{n-1}\del_m\wt{y}_m^{(i,j)}(t)\right)
\eea
to be a $\mbb{Q}^{(i)}_B$-martingale within an each interval $(T_{q(t)-1},T_{q(t)})$. 
Here, we have put
\bea
\wt{y}_m^{(i,j)}(t):=c_m^{(i)}(t)+y_m^{(i,j)}(t)
\eea
for notational simplicity.

By following the same arguments given in Proposition~\ref{prop-cdisc}, the 
no-arbitrage dynamics of $\wt{y}_{n}^{(i,j)}$ for $n\in\{0,\cdots,N^*-1\}$ is give by
\bea
d\wt{y}_n^{(i,j)}(t)=\left\{ \wt{\sigma}_n^{(i,j)}(t)\cdot \Bigl(\sum_{m=q(t)}^{n-1}\del_m
\wt{\sigma}_m^{(i,j)}(t)\Bigr)+\frac{1}{2}\del_n|\wt{\sigma}_n^{(i,j)}|^2\right\}dt+
\wt{\sigma}_n^{(i,j)}(t)\cdot dW_t^{\mbb{Q}^{(i)}_B}
\eea
with some appropriate $\mbb{F}$-adapted processes 
$\wt{\sigma}_m^{(i,j)}:\Omega\times [0,T_m]\rightarrow \mbb{R}^d$, $m\in\{0,\cdots,N^{*}-1\}$.
Thus, by the relation $dy^{(i,j)}_n(t)=d\wt{y}_n^{(i,j)}(t)-dc_n^{(i)}(t)$, 
the drift term is obtained by
\bea
\alpha_n^{(i,j)}(t)&=&\wt{\sigma}_n^{(i,j)}(t)\cdot \Bigl(\sum_{m=q(t)}^{n-1}\del_m
\wt{\sigma}_m^{(i,j)}(t)\Bigr)+\frac{1}{2}\del_n|\wt{\sigma}_n^{(i,j)}|^2\nn \\
&&-\sigma_n^{(i)}(t)\cdot \left(\sum_{m=q(t)}^{n-1}\del_m \sigma_m^{(i)}(t)\right)-
\frac{1}{2}\del_n |\sigma_n^{(i)}(t)|^2~.
\eea
By the definition of $\wt{y}^{(i,j)}$, we have for every $m\in\{0,\cdots,N^{*}-1\}$,
\bea
\wt{\sigma}_m^{(i,j)}(t)=\sigma_m^{(i)}(t)+\sigma_{y,m}^{(i,j)}(t)~,
\eea
and hence
\bea
\alpha_n^{(i,j)}(t)&=&\sigma_{y,n}^{(i,j)}(t)\cdot \Bigl(\sum_{m=q(t)}^{n-1}\del_m 
[\sigma_{y,m}^{(i,j)}(t)+\sigma_m^{(i)}(t)]\Bigr)+\sigma_n^{(i)}(t)\cdot \Bigl(\sum_{m=q(t)}^{n-1}\del_m \sigma_{y,m}^{(i,j)}(t)\Bigr)\nn \\
&&+\frac{1}{2}\del_n |\sigma_{y,n}^{(i,j)}(t)|^2+\del_n (\sigma_{y,n}^{(i,j)}(t)\cdot \sigma_n^{(i)}(t))
\eea
which proves the claim.
\end{proof}
\end{proposition}

\subsubsection*{Remark}
Although the dynamics of the funding spread is rather complicated, it has exactly 
the same structure as the forward {\it hazard} rate model~\cite{Schonbucher}.
Thus, if a term structure model of a 
default intensity has  already been implemented, then modifying it for the stochastic funding spread 
should not be difficult for practitioners.

\subsection{Dynamics of foreign exchange rates}
Finally, let us consider the dynamics of the foreign exchange rate
and the associated measure change.
The spot foreign exchange rate $\{f_x^{(i,j)}(t)\}_{t\geq 0}$ itself is not tradable
but the foreign currency is.

We can define the measure change between $\mbb{Q}^{(i)}_B$ and $\mbb{Q}^{(j)}_B$ by
\bea
\frac{d\mbb{Q}^{(j)}_B}{d\mbb{Q}_B^{(i)}}\Bigr|_{\calf_t}=\frac{B^{(j)}(t)f_x^{(i,j)}(t)}{B^{(i)}(t)f_x^{(i,j)}(0)}
\eea
since the right hand side should be a positive $\mbb{Q}_B^{(i)}$-martingale for the 
absence of arbitrage.
This fact immediately tells us that
\begin{proposition}
The no-arbitrage dynamics of the foreign exchange rate of a foreign currency $(j)$  
with respect to a domestic currency $(i)$ is given by
\bea
\label{eq-spot-fx}
df_x^{(i,j)}(t)/f_x^{(i,j)}(t)&=&\bigl(r^{(i)}(t)-r^{(j)}(t)\bigr)dt+\sigma_X^{(i,j)}(t)\cdot dW_t^{\mbb{Q}_B^{(i)}}\nn \\
&=&\bigl(c^{(i)}(t)-c^{(j)}(t)+y^{(i,j)}(t)\bigr)dt+\sigma_X^{(i,j)}(t)\cdot dW_t^{\mbb{Q}_B^{(i)}}
\eea
with some appropriate volatility process $\sigma_X^{(i,j)}:\Omega\times [0,T_{N^*}]\rightarrow \mbb{R}^d$,
which is $\mbb{F}$-adapted.
\end{proposition}

By the non-randomness assumption on the shortest bonds, one can easily check that the relation
\bea
&&\frac{d\mbb{Q}^{(j)}_B}{d\mbb{Q}_B^{(i)}}\Bigr|_{\calf_t}=\frac{C^{(j,k)}(t)f_x^{(i,j)}(t)}{C^{(i,k)}(t)f_x^{(i,j)}(0)} 
\eea
holds for an arbitrary currency $(k)$. 
It is interesting to observe that one must use the common collateral currency 
to define the measure change. This is due to 
the existence of the funding spreads $\{y^{(i,j)}\}$,
which is a direct consequence of the non-zero cross currency basis being observed in the market~\cite{ccc-orig}.

By arranging the right hand side
and using the definition (\ref{eq-fc-fx}),
one sees that
\be
f_x^{(i,j)}(t,T_{q(t)};(k)),t\in[T_{q(t)-1},T_q(t)]
\ee
is a $\mbb{Q}^{(i)}_{B}$-martingale regardless of the collateral currency $(k)$.
In addition, its volatility must be identical to that of the spot exchange rate
so that its gives the same Radon-Nikodym density.
This gives the following result:
\begin{proposition}
The no-arbitrage dynamics of the rolling forward exchange rate of 
a foreign currency $(j)$ with respect to a domestic currency $(i)$ collateralized by a currency $(k)$ is given by
\bea
df_x^{(i,j)}(t,T_{q(t)};(k))=f_x^{(i,j)}(t,T_{q(t)};(k))\sigma_X^{(i,j)}(t)\cdot dW_t^{\mbb{Q}^{(i)}_B}~,
\eea
with a condition at each rollover date $f_x^{(i,j)}(T_{n},T_{n+1};(k))=f_x^{(i,j)}(T_n,T_n;(k))
\displaystyle\frac{\wt{Y}^{(j,k)}(T_n,T_{n+1})}{\wt{Y}^{(i,k)}(T_n,T_{n+1})}$
for every $n\in\{0,\cdots,N^*-1\}$.
$\sigma_X^{(i,j)}:\Omega\times [0,T_{N^*}]\rightarrow \mbb{R}^d$ is an $\mbb{F}$-adapted volatility process
equal to that of the spot exchange rate~(\ref{eq-spot-fx}).
\end{proposition}
Note that $f_x^{(i,j)}(t,t;(k))=f_x^{(i,j)}(t)$.
The following corollary is obvious by the standard application of Girsanov-Maruyama theorem.
\begin{corollary}
The Brownian motion under the measure $\mbb{Q}^{(j)}_B$ is related to that of 
$\mbb{Q}^{(i)}_B$ by
\bea
W_t^{\mbb{Q}_B^{(j)}}=W_t^{\mbb{Q}_B^{(i)}}-\int_0^t \sigma_X^{(i,j)}(s)ds~.
\eea
\end{corollary}

\subsection{Dynamics of Equities}
If one tries to model a spot equity dynamics directly, it requires to 
simulate the risk-free rate $r$ or the dividend yield implied by the collateral account $y$, 
neither of them is easy to observe. The best way to avoid this problem is 
to use the equity forward price.
Let us consider a price process of a equity $\{S_t^{(i)}\}_{t\geq 0}$ denominated
by currency $(i)$. We do not ask whether the equity has dividend payments or not.
Let us denote the price of a forward contract on this equity with maturity $T_n$
collateralized by the same currency $(i)$ as
$S^{(i)}(t,T_n)$:
\be
S^{(i)}(t,T_n)=\frac{\mbb{E}_t^{\mbb{Q}^{(i)}_B}\Bigl[e^{-\int_t^{T_n} c^{(i)}(s)ds}S^{(i)}(T_n)\Bigr]}{D^{(i)}(t,T_n)},
\quad t\in[0,T_n].
\ee
Then, by exactly the same reasoning of Proposition~\ref{prop-lois}, one obtains:
\begin{proposition}
The no-arbitrage dynamics of the equity forward price is given by 
\bea
dS^{(i)}(t,T_n)=S^{(i)}(t,T_n)\left(\sigma_{S^{(i)},n}(t)\cdot \Bigl(\sum_{m=q(t)}^{n-1}\del_m \sigma_m^{(i)}(t)\Bigr)dt
+\sigma_{S^{(i)},n}(t)\cdot dW_t^{\mbb{Q}^{(i)}_B}\right)
\eea
for $t\in[0,T_n],~n\in\{0,\cdots,N^{*}-1\}$ with some appropriate $\mbb{F}$-adapted volatility process $\sigma_{S^{(i)},n}:\Omega\times[0,T_{n}]\rightarrow\mbb{R}^d$.
\end{proposition}
The same technique can be applied to an arbitrary asset or index, as long as there exists 
a corresponding forward market.

\subsection{Summary of the dynamics}
For easy reference, we summarize the dynamics of various rates 
in the multi-currency setup
with a base currency $(i)$ below. For actual implementation,
one has to decide the size of dimension ``$d$" and reproduce the 
observed correlation among the underlyings as accurately as possible.
For these points, we recommend Rebonato (2004)~\cite{Rebonato}),
which contains many valuable insights for implementation, in particular Chapter 19 and 20.
\\
{\bf{Rates for the base currency} }
\bea
&&dc_n^{(i)}(t)=\left\{\sigma_n^{(i)}(t)\cdot \left(\sum_{m=q(t)}^{n-1}\del_m \sigma_m^{(i)}(t)\right)+
\frac{1}{2}\del_n |\sigma_n^{(i)}(t)|^2 \right\}dt+\sigma_n^{(i)}(t)\cdot dW_t^{\mbb{Q}_B^{(i)}}~ \\
&&\frac{dB^{(i)}(t;T_{n-1},T_n)}{B^{(i)}(t;T_{n-1},T_n)}=
\sigma_{B,n}^{(i)}(t)\cdot\Bigl(\sum_{m=q(t)}^{n-1}\del_m \sigma_m^{(i)}(t)\Bigr)dt+\sigma_{B,n}^{(i)}(t)
\cdot dW_t^{\mbb{Q}_B^{(i)}}
\eea
{\bf{Funding spreads with respect to the base currency}}
\bea
&&dy_n^{(i,j)}(t)=\left\{\frac{\bigl.}{\bigr.} \right. \sigma_{y,n}^{(i,j)}(t)\cdot \Bigl(\sum_{m=q(t)}^{n-1}\del_m 
[\sigma_{y,m}^{(i,j)}(t)+\sigma_m^{(i)}(t)]\Bigr)+\sigma_n^{(i)}(t)\cdot \Bigl(\sum_{m=q(t)}^{n-1}\del_m \sigma_{y,m}^{(i,j)}(t)\Bigr)\nn \\
&&\qquad\left. 
+\frac{1}{2}\del_n |\sigma_{y,n}^{(i,j)}(t)|^2+\del_n (\sigma_{y,n}^{(i,j)}(t)\cdot \sigma_n^{(i)}(t))\frac{\bigl.}{\bigr.}\right\}dt+\sigma_{y,n}^{(i,j)}(t)\cdot dW_t^{\mbb{Q}_B^{(i)}}~
\eea
{\bf{Foreign exchange rate}}
\bea
df_x^{(i,j)}(t)/f_x^{(i,j)}(t)=\Bigl(c^{(i)}_t-c^{(j)}_t+y^{(i,j)}_t\Bigr)dt+\sigma_X^{(i,j)}(t)\cdot dW_t^{\mbb{Q}^{(i)}}
\eea
{\bf{Rates for a foreign currency}}
\bea
&&dc_n^{(j)}(t)=\left\{\sigma_n^{(j)}(t)\cdot \left(\sum_{m=q(t)}^{n-1}\del_m \sigma_m^{(j)}(t)-\sigma_X^{(i,j)}(t)\right)+
\frac{1}{2}\del_n |\sigma_n^{(j)}(t)|^2 \right\}dt+\sigma_n^{(j)}(t)\cdot dW_t^{\mbb{Q}_B^{(i)}}~\nn \\
&&\frac{dB^{(j)}(t;T_{n-1},T_n)}{B^{(j)}(t;T_{n-1},T_n)}=
\sigma_{B,n}^{(j)}(t)\cdot\Bigl(\sum_{m=q(t)}^{n-1}\del_m \sigma_m^{(j)}(t)-\sigma_X^{(i,j)}(t)\Bigr)dt+\sigma_{B,n}^{(j)}(t)
\cdot dW_t^{\mbb{Q}_B^{(i)}}
\eea
{\bf{Funding spreads with respect to a foreign currency}}
\bea
&&\hspace{-5mm}dy_n^{(j,k)}(t)=\left\{\frac{\bigl.}{\bigr.} \right. \sigma_{y,n}^{(j,k)}(t)\cdot \Bigl(\sum_{m=q(t)}^{n-1}\del_m 
[\sigma_{y,m}^{(j,k)}(t)+\sigma_m^{(j)}(t)]-\sigma_X^{(i,j)}(t)\Bigr)+\sigma_n^{(j)}(t)\cdot \Bigl(\sum_{m=q(t)}^{n-1}\del_m \sigma_{y,m}^{(j,k)}(t)\Bigr)\nn \\
&&\qquad\left. 
+\frac{1}{2}\del_n |\sigma_{y,n}^{(j,k)}(t)|^2+\del_n (\sigma_{y,n}^{(j,k)}(t)\cdot \sigma_n^{(j)}(t))\frac{\bigl.}{\bigr.}\right\}dt+\sigma_{y,n}^{(j,k)}(t)\cdot dW_t^{\mbb{Q}_B^{(i)}}~
\eea

\section{Remarks on a risk-free money market account}
\label{sec-risk-free}
It is clear from the results in Sections \ref{sec-HJM}
and \ref{sec-disc-HJM}, one need not refer to {\it risk-free interest rates}
and the associated money-market accounts in a fully collateralized market. 
In fact, it is matter of preference for the user to  
choose a certain currency $(i)$ as a base currency and treat its collateral account $C^{(i)}$
as a unique risk (default)-free bank account. 
It makes $y^{(i)}=0$ but does not change dynamics given in the previous sections 
at all.  

However, due to the presence of non-zero cross currency basis, this requires
asymmetric treatments for the other collateral rates of foreign currencies,
which cannot be treated as traditional risk-free rates.
Furthermore, it has become clear that the authority can force the collateral rate to be negative 
as observed in recent EONIA market, for example.
Considering the fact that there exist firms which are outside the banking regulation 
and do not have to collateralize their contracts, 
it is not always rational to treat the overnight rate as risk-free. 
Even for those who are forced to collateralize the contracts,
it is more natural to consider the collateral agreement has
some dividend yield which makes overnight rate effectively negative.

In our setup, there is no problem to make a collateral rate negative since there is no
a priori restriction to the dividend yield process $\{y^{(i)}\}$. 
Due to these observations, we have chosen to 
use a risk-free money market account separately from a collateral account.
The biggest assumption for our framework is the existence of a common ``risk-free" money market 
account for every market participant.
It is completely rational that each financial firm
wants to reflect its own funding/investment conditions instead of a common ``risk-free" rate,
but then it will inevitably produce the company dependent price of derivatives~\cite{Crepey-book}.
Therefore, we think that the company-specific effects should be treated in FVA, 
separately from the benchmark pricing we have described in this article.

\section{Conclusion}
This paper is an extension of previous works~\cite{dynamic_basis, ccc-old, ccc-orig}
and provided more detailed explanation for the general framework for
the interest rate modeling in a fully collateralized market.
In particular, we gave a new formulation for the funding spread
dynamics which is more suitable in the presence of non-zero correlation 
to the collateral rates.
We also presented a complete picture 
of a discretized HJM model with a fixed tenor structure under a multi-currency setup, 
which is arbitrage free, readily 
implementable and capable of taking the stochastic cross currency basis into account.



\begin{thebibliography}{99}

\bibitem{bianchetti}
Bianchetti, M (2010) Two curves, one price,
 {\it Risk Magazine, Issue August},  74-80.

\bibitem{Bianchetti-Riskbook}
Bianchetti, M and M Morini (edit) (2013),  Interest Rate Modeling after the Financial Crisis,
{\it Risk Books, London.}

\bibitem{BGM}
Brace, A, D Gatarek and M Musiela (1997), The market model of interest rate dynamics,
{\it 1997, Math. Finance}, $\bf{7}$(2), 127-155.


\bibitem{Brigo-book}
Brigo, D, M Morini and A Pallavicini (2013), Counterparty Credit Risk,
Collateral and Funding,
{\it Wiley, West Sussex.}

\bibitem{Crepey-Grbac}
Cr\'epey, S, Z Grbac and H Nguyen (2012),  A multiple-curve HJM model
of interbank risk, {\it Math Finan Econ}, 6, 155-190.

\bibitem{Crepey-book}
Cr\'epey, S, T Bielecki with an introductory dialogue by D Brigo (2014),
Counterparty Risk and Funding, {\it CRC press, NY.}


\bibitem{trolle}
Filipovi\'c, D and A Trolle (2013),  The term structure of interbank risk,
{\it Journal of Financial Economics}, 109, 707-733.

\bibitem{note-curve}
Fujii, M, Y  Shimada and A Takahashi (2010a), A Note on Construction of Multiple Swap Curves with and without Collateral,
{\it FSA Research Review}, Vol.6, 139-157.

\bibitem{ccc-old}
Fujii, M, Y Shimada and A Takahashi (2010b), Collateral posting and choice of collateral currency,
{\it available at SSRN: 1601866.}

\bibitem{dynamic_basis}
Fujii, M, Y Shimada and A Takahashi (2011), A Market Model of Interest Rates with 
Dynamic Basis Spreads in the presence of Collateral and Multiple Currencies,
{\it Wilmott Magazine}, Issue 54: 61-73.

\bibitem{ccc-orig}
Fujii, M and A Takahashi (2011), Choice of Collateral Currency,
{\it Risk Magazine}, January Issue, 120-125.

\bibitem{imperfect_collateral}
Fujii, M and A Takahashi (2013a),  Derivative Pricing under Asymmetric and Imperfect Collateralization, and CVA,
{\it Quantitative Finance}, Vol. 13, Issue 5, 749-768.

\bibitem{RiskBook}
Fujii, M and A Takahashi, (2013b), Interest Rate Modeling under full Collateralization, in
M Morini and M Bianchetti (editors), {\it Interest Rate Modeling after the Financial Crisis},  
pp. 241-282,
Risk Books, London.


\bibitem{Grbac-Papapantoleon}
Grbac, Z, A Papapantoleon, J Schoenmakers and D Skovmand (2014), Affine Libor models
with multiple curves,
{\it preprint available in arXiv.}

\bibitem{HJM}
Heath, D, R Jarrow and A Morton (1992), Bond pricing and the term structure of interest rates, {\it Econometrica}, 
60(1), 77-105.

\bibitem{ISDA}
ISDA Margin Survey 2011.


\bibitem{Mercurio}
Mercurio, F (2009), Interest rate and the credit crunch: New formulas and market models,
{\it Bloomberg Portfolio Research Paper.}

\bibitem{piterbarg}
Piterbarg, V (2010), Funding beyond discounting : collateral agreements and derivatives pricing,
{\it Risk Magazine}, Issue February, 97-102.

\bibitem{Rebonato}
Rebonato, R (2004), Volatility and correlation, 
{\it Wiley, West Sussex.}

\bibitem{FXLMM}
Schl\"ogl, E  (2002),  A multi-currency extension of the lognormal interest rate market models,
{\it Finance and Stochastics}, 6, 173-196.

\bibitem{Schonbucher}
Sch\"onbucher, P (2003),  Credit Derivatives Pricing Models: models, pricing and implementation,
{\it Wiley, UK}.

\bibitem{T}
Takahashi, A (2015),
Asymptotic Expansion Approach in Finance, in P Friz, J Gatheral, A Gulisashvili, A Jacquier
and J Teichmann (editors), {\it Large Deviations and Asymptotic Methods in Finance},
pp. 345-411, Springer.


\end{thebibliography}
\end{document}